\documentclass[aps,prl,twocolumn,amsmath,amssymb,superscriptaddress]{revtex4-1}

\usepackage{graphicx}
\usepackage{bm}
\usepackage{amsmath}
\usepackage{verbatim}     
\usepackage[dvipsnames]{xcolor}       
\usepackage[normalem]{ulem}   

\newcommand{\ket}[1]{|#1\rangle}                      
\newcommand{\hg}[1]{\hbox{HyGG}_{#1}}		

\usepackage{hyperref}
\hypersetup{%
   pdfpagemode=None, 
   pdfstartpage=1,
   pdfmenubar=true,
   pdftoolbar=true,
   colorlinks = true,
   linkcolor=blue,
   citecolor=blue,
   urlcolor=blue,
   bookmarksopen=false
 }

\begin{document}
\title{Topological features of vector vortex beams perturbed with uniformly polarized light}
\author{Alessio D'Errico}
\affiliation{Dipartimento di Fisica, Universit\`{a} di Napoli Federico II, Complesso Universitario di Monte Sant'Angelo, Napoli, Italy}
\author{Maria Maffei}\altaffiliation{currently also at: ICFO-Institut de Ciencies Fotoniques, The Barcelona Institute of Science and Technology, 08860 Castelldefels, Spain}
\affiliation{Dipartimento di Fisica, Universit\`{a} di Napoli Federico II, Complesso Universitario di Monte Sant'Angelo, Napoli, Italy}
\author{Bruno Piccirillo}
\affiliation{Dipartimento di Fisica, Universit\`{a} di Napoli Federico II, Complesso Universitario di Monte Sant'Angelo, Napoli, Italy}
\author{Corrado de Lisio}
\affiliation{Dipartimento di Fisica, Universit\`{a} di Napoli Federico II, Complesso Universitario di Monte Sant'Angelo, Napoli, Italy}
\affiliation{CNR-SPIN, Complesso Universitario di Monte Sant'Angelo, Napoli, Italy}
\author{Filippo Cardano}\email{correspondence to: filippocardano@gmail.com}
\affiliation{Dipartimento di Fisica, Universit\`{a} di Napoli Federico II, Complesso Universitario di Monte Sant'Angelo, Napoli, Italy}
\author{Lorenzo Marrucci}
\affiliation{Dipartimento di Fisica, Universit\`{a} di Napoli Federico II, Complesso Universitario di Monte Sant'Angelo, Napoli, Italy}
\affiliation{CNR-ISASI, Institute of Applied Science and Intelligent Systems, Via Campi Flegrei 34, Pozzuoli (NA), Italy}
%

\begin{abstract} 
Optical singularities manifesting at the center of vector vortex beams are unstable, since their topological charge is higher than the lowest value permitted by Maxwell's equations. Inspired by conceptually similar phenomena occurring in the polarization pattern characterizing the skylight, we show how perturbations that break the symmetry of radially symmetric vector beams lead to the formation of a pair of fundamental and stable singularities, i.e. points of circular polarization. We prepare a superposition of a radial (or azimuthal) vector beam and a uniformly linearly polarized Gaussian beam; by varying the amplitudes of the two fields, we control the formation of pairs of these singular points and their spatial separation. We complete this study by applying the same analysis to vector vortex beams with higher topological charges, and by investigating the features that arise when increasing the intensity of the Gaussian term. Our results can find application in the context of singularimetry, where weak fields are measured by considering them as perturbations of unstable optical beams. 
\end{abstract}

\maketitle
\onecolumngrid
\section*{Introduction}\label{sec:main}
Light beams showing an inhomogeneous polarization distribution, commonly referred to as vector beams (VBs), represent a precious resource in an increasing number of photonic applications \cite{Cardano2015}: astronomy \cite{Mawet2010}, microscopy \cite{Dorn2003,Abouraddy2006}, optomechanics \cite{Shvedov2014,Rui2016}, materials structuring \cite{Anoop2014},  nanophotonics \cite{Li2012,Neugebauer2014} and quantum sciences \cite{DAmbrosio2012,Fickler2014,Aiello2014,Cardano2016} are some remarkable examples. Uniformly polarized beams can be easily converted into such spatially structured fields by coupling the vectorial and the spatial degrees of freedom of light \cite{Cardano2015,Bliokh2015}, as recently demonstrated in a variety of photonic architectures \cite{Oron2000,Niv2005,Maurer2007,Beckley2010,Cardano2012,Gong2014,Bouchard2014,Naidoo2015,Chille2016,Radwell2016}. The fine structure of VBs polarization may show several typologies of singular points \cite{Nye1987,Berry2001,Freund2002a,Soskin2003,Dennis2009}, in close analogy to other inhomogeneous systems (fingerprint, tidal heights across the oceans, etc.). 
Here we consider those spatial regions where there is no preferred direction for the oscillations of the electric field \cite{Nye1987,Berry2001,Dennis2009}, with the most relevant case being represented by the so called $C$-points, that is points where the polarization is circular. Their formation and dynamical evolution have been investigated in the complex polarization pattern characterizing several structured fields, such as for instance speckle fields \cite{Flossmann2008}, random superposition of vector waves \cite{Berry2001}, light passing through inhomogeneous anisotropic media \cite{Beckley2010,Cardano2013}, photonic crystals \cite{Burresi2009,Lang2015}. Independently of the specific system, the electric field around a polarization singularity is oriented according to the value of the associated topological charge $\eta$; this is an integer or semi-integer number, defined as the angle described by the major axis of the local polarization ellipse (divided by 2$\pi$) when following a closed path around the $C$-point. Besides its connection with the surrounding polarization distribution \cite{Freund2001}, the value of this charge is particularly important in determining the singularity robustness, since only the lowest order $C$-points with $\eta=\pm 1/2$ are stable with respect to small deformations of the optical system \cite{Lang2015,Bliokh2008,Freund2002,Dennis2002,Nye1987}. This is analogous to the case of high-order optical vortices in scalar fields, which have been observed to split into elementary vortices as soon as a tiny perturbation is introduced \cite{Bekshaev2004,Dennis2006,Freund1999,Kumar2011,Ricci2012,Soskin1997}. The instability of higher-order polarization singularities, with the role of $C$-points played by points of unpolarized light, can be beautifully observed in the skylight polarization pattern, where, differently from the case of $C$-points, such singularities are loci where the light is fully unpolarized; in the sky, the original two singular points (for which $\eta$ = 1), positioned at the Sun and the anti-Sun loci, split into four slightly displaced lowest-order singularities (with $\eta$ = 1/2) because of the contribution from multiple Rayleigh scattering of sunlight in the atmosphere \cite{Berry2004}. Inspired by these phenomena, but in the context of fully polarized laser light, here we investigate the formation of lowest order $C$-points at the center of vector vortex beams (VVBs) \cite{Niv2006,Holleczek2010} in presence of a weak perturbing field.
VVBs are a particular class of vector beams (radially and azimuthally polarized beams are remarkable examples) for which a polarization singularity and an optical vortex (phase singularity) are superimposed at the center of the beam. This happens because the spatial mode associated with each of the two opposite circular polarizations is that of a light helical mode. These peculiar optical spatial modes, described in terms of an integer number $m$, show a helical wavefront and carry a definite amount of orbital angular momentum (OAM) \cite{Yao2011}, equal to $m\hbar$ per photon. The field amplitude vanishes at the center of these beams, where the associated phase is not defined and an optical vortex with topological charge $m$ appears. Since left and right circular components are both helical modes, in VVBs the total field is vanishing along the beam axis and its orientation is undefined. This peculiar polarization singularity is typically referred to as $V$-point \cite{Freund2002}; unlike the case of a $C$-point, here the instantaneous oscillation direction of the electric field is undefined (at any time). The lowest topological charge admitted for $V$-points is $\pm1$, since these are singularities of a field of vectors (the instantaneous electric field), whereas $C$-points refer to a field of ellipses (the trajectory described by the vector in a temporal cycle). Here we show that a small perturbation changes the nature of the vector field characterizing pure VVBs, whose local polarization states acquire a tiny ellipticity. Since in such a field Maxwell's equations allows for polarization singularities with a lower charge ($C$-points), even the lowest order $V$-point becomes unstable and unfolds into a pair of equally charged $C$-points \cite{Freund2002}. We investigate experimentally this mechanism by perturbing a radial and azimuthal VVB ($\eta=1$) with a uniformly polarized beam, and complete the analysis with an example of higher order VVB ($\eta=2$). This kind of perturbation acts as a coherent background, whose role has been investigated in the decay of optical vortices at the center of beams carrying OAM \cite{Ricci2012}. Recently a similar study of $V$-point unfolding was proposed theoretically in Ref.\ \cite{Vyas}, although the analysis is focused on $V$-point and $C$-point dynamics during the beam propagation, rather than the instability of $V$-points. By controlling the amplitude of the two fields, we report the progressive formation of $C$-points (that originate from the central $V$-point), whose separation increases as the Gaussian term gets higher. Importantly, the polarization pattern modification is always accompanied by a deformation of the original intensity pattern. Interesting features arise when increasing the intensity of the perturbing term, in particular when this becomes equal or higher than the original VVB and the polarization pattern may lose its non-trivial topological features (at least in the region where almost all the field energy is enclosed).
\begin{figure}[t!]
\centering
\includegraphics[width=\textwidth]{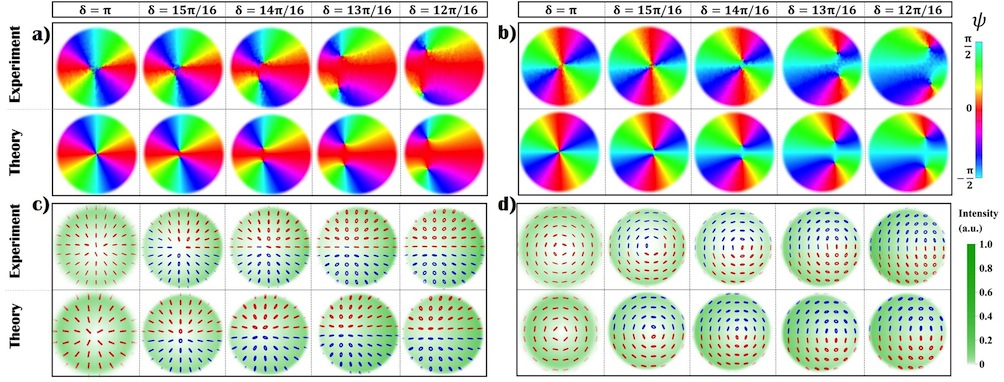}
\caption{Instability of polarization singularities at the center of a VVB. The instability of a $V$-point with topological charge $\eta=1$ is investigated experimentally by changing the $q$-plate retardation with respect to the optimal condition $\delta=\pi$. In panels a-b we plot the quantity $\psi=\frac{1}{2}Arg(S_1+i\,S_2)$, that represents the orientation of polarization ellipses, for a radial (panel a) and an azimuthal (panel b) VVB. Colors associated with different values of $\psi$ are shown in the figure legend.  As visible in the figure labels, plots are obtained when varying $\delta$ in the range $\{12\pi/16,\pi\}$ with steps of $\pi/16$. Importantly, $C$-points and $V$-points appear here as vortices of the complex scalar field $S_1+i\,S_2$; as we decrease $\delta$, two $C$-points clearly appear in the place of the $V$-point. As discussed in the main text, we are considering only a small area (of the order of $w^2/4$, where $w$ is the beam radius) at the center of the beam, where the singularity transformation is taking place. In c-d, we show the corresponding experimental and theoretical polarization patterns associated with these fields. Here, red and blue colored  ellipses are associated with left ($s_3>1$) and right handed ($s_3<1$) polarization states, respectively.}
\label{fig: V-stability}
\end{figure}

\section*{Results}\label{sec:main}
{\bf $V$-point instability at the center of a vector vortex beam}
A VVB corresponds to the superposition of two (or more) different helical modes of light associated with orthogonal circular polarizations. Denoting as $\ket{L}$ and $\ket{R}$ states of left and right circular polarizations, respectively, a VVB can be written as
\begin{align}\label{eq:vvb}
\ket{VVB}_m=f(r,z)\left(c_L \ket{L} \,e^{-i\,m\,\phi}+c_R \ket{R}\,e^{i\,m\,\phi}\right),
\end{align}
where $c_L\neq0$, $c_R\neq0$ are complex coefficients and the complex amplitude $f(r,z)\,e^{i\,m\,\phi}$ describes the field associated with a helical mode of order $m$, expressed in terms of the cylindrical coordinates $(r,\phi,z)$ with the $z$ axis corresponding to the optical axis of the beam. While the phase factor $e^{i\,m\,\phi}$ gives rise to the typical helical wavefront, the function $f(r,z)$ describes the radial distribution of the field. Here we have considered the simple case of two helical modes with opposite $m$; similar results hold for all other VVBs. Except the case $m=0$, helical modes vanish along the optical axis ($r=0$) where they show an optical vortex with charge $m$ \cite{Soskin1997}. Accordingly, in the case of an ideal VVB, a $V$-point with charge $\eta=m$ is present at the center of the beam, at any plane transverse to the propagation direction. In order to show a possible mechanism that leads to the $V$-point transformation into lowest order singularities, let us consider the specific case $m=1$ and $c_L=c_R=1$, corresponding to a radially polarized beam. At a fixed transverse plane and very close to the beam center, that is at $r$ much smaller than the typical beam dimensions, Eq.\ \ref{eq:vvb} has a simpler expression:
\begin{align}\label{eq:vvb_smallr}
\ket{VVB}_1^{rad}\simeq A\,r\left(\ket{L} \,\,e^{-i\,\phi}+\ket{R}\,e^{i\,\phi}\right),
\end{align}
where $A$ is a real constant defining the field intensity. We add to Eq.\ \ref{eq:vvb_smallr} a linearly polarized term with uniform amplitude $\epsilon\,e^{i\,\alpha_\epsilon}$ ($\epsilon$ and $\alpha_\epsilon$ are real constants), whose electric field is oriented at an angle $\theta$ with respect to the horizontal direction. In the representation of circular polarizations, this perturbation can be written as $\ket{\epsilon,\theta}=\epsilon \,e^{i\,\alpha_\epsilon}\left(e^{-i\,\theta}\,\ket{L}+e^{\,i\,\theta}\,\ket{R}\right)$; when added to the original VVB, Eq.\ \ref{eq:vvb_smallr} becomes
\begin{align}\label{eq:vvb_perturb}
\ket{VVB}_1^{rad} &\rightarrow \ket{VVB}_1^{rad}+\ket{\epsilon,\theta}\simeq\nonumber\\&\simeq(A\,r\,e^{-i\,\phi}+\epsilon\,e^{i\,(\alpha_\epsilon-\theta)})\ket{L} +(A\,r\,e^{i\,\phi}+\epsilon\,e^{i\,(\alpha_\epsilon+\theta)})\ket{R}.
\end{align} 
Left and right $C$-points are located at positions $(r_L,\phi_L)$ and $(r_R,\phi_R)$ where the right and left circular components of the field are vanishing, respectively. It is straightforward to see that
\begin{align}\label{eq:cpoints}
r_L=r_R=\epsilon/A; \quad \phi_R=\theta-\alpha_\epsilon-\pi; \quad \phi_L=\theta+\alpha_\epsilon+\pi.
\end{align} 
In general $\phi_L\neq\phi_R$ and two $C$-points with opposite handedness but with equal charge $\eta=1/2$ generate from the original $V$-point. Using the same approach it is possible to show that $V$-points of order $m$ unfold into $2m$ $C$-points, with the sign of their charge being equal to that of the original singularity.\\
\linebreak
{\bf Generation and perturbation of a VVB using electrically tunable q-plates}
The results we discussed in the previous section can be easily simulated experimentally by exploiting the same approach reported in \cite{Cardano2012} for the generation of pure VVBs. The preparation and the controlled alteration of a VVB is obtained by tuning the spin-orbit interaction of a light beam in a $q$-plate \cite{Marrucci2006,Marrucci2011}. This is an optical element made of a thin layer of liquid crystals, whose optic axes are arranged so as to form a singular inhomogeneous pattern. Besides the topological charge $q$, defining the rotation of the local liquid crystal axis around the singular point (divided by $2\pi$), the action of this device is determined by its optical retardation $\delta$. The value of the latter can be suitably adjusted by applying an external electric field, which allows for controlling the strength of the spin-orbit interaction mediated by the plate \cite{Piccirillo2010}. When passing through a $q$-plate placed at the beam waist (this is the standard configuration that we adopt throughout the manuscript), a TEM$_{0,0}$ Gaussian beam with uniform left or right circular polarization is transformed as follows \cite{Karimi2009}:
\begin{align}\label{eq:qplate_action}
\hat Q_{\delta}\, \hbox{TEM}_{0,0}\,\ket{L}&=\cos{\left(\delta/2\right)}\hbox{TEM}_{0,0}\,\ket{L}+i \sin{\left(\delta/2\right)}\hg{-|2q|,|2q|}\,e^{i\,2(q\,\phi+\alpha_0)}\,\ket{R},\nonumber\\
\hat Q_{\delta}\, \hbox{TEM}_{0,0}\,\ket{R}&=\cos{\left(\delta/2\right)}\hbox{TEM}_{0,0}\,\ket{R}+i \sin{\left(\delta/2\right)}\hg{-|2q|,|2q|}\,e^{-i\,2(q\,\phi+\alpha_0)}\,\ket{L},
\end{align}
 where $\alpha_0$ is the angle of the liquid crystals optic axis at $\phi=0$.
The previous equation shows that left and right circular polarizations are partially converted into helical modes of order $\pm 2q$, respectively, with the amount of converted light depending on the value of $\delta$ ($\delta=\pi$ corresponds to a full conversion). Helical modes generated by a $q$-plate are described by the so-called HyperGeometric-Gaussian modes (HyGG$_{p,m}$, which in our notation corresponds to the radial amplitude profile only of the mode, as the azimuthal phase factor is written explicitly) \cite{Karimi2007}, corresponding to a specific class of light beams carrying OAM \cite{Vallone2015}, analogously to Laguerre-Gauss or Bessel beams. Two indices $(p,m)$ specify the mode properties, where $m$ is associated with the OAM content while $p$ determines the radial distribution of the field. It is worth noting that the same helical modes with HyGG radial structure are also generated by any optical device (spiral phase plates, pitchfork holograms, etc.) that suddenly imprints an azimuthal phase factor on the input field, with such device placed in the focal region of the beam. Hence, besides our specific setup, our analysis applies to all configurations in which VVBs are generated relying on this approach. Eq.\ \ref{eq:qplate_action} clearly shows that, if $\delta=\pi$, a linearly polarized Gaussian beam is fully converted into a VVB, showing a $V$-point with charge $\eta=2q$ at its center. In particular, azimuthally and radially polarized beams are obtained when the input polarization is vertical $(V)$ and horizontal $(H)$, respectively, the plate charge is $q=1/2$ and $\alpha_0=0$ (see Fig.\ \ref{fig: V-stability}):
\begin{align}\label{eq:qplate_action_pi}
\hat Q_{\pi}\, \hbox{TEM}_{0,0}\,\ket{H}&=i\,\ket{VVB}_1^{rad},\nonumber\\
\hat Q_{\pi}\, \hbox{TEM}_{0,0}\,\ket{V}&=i\,\ket{VVB}_1^{az}.
\end{align} 
where kets $\ket{H}$ and $\ket{V}$ represent $H$ and $V$ polarizations states.
When changing the value of the retardation to $\delta=\pi+2\varepsilon$, with $\varepsilon\ll\pi$, a fraction $\varepsilon$ of the input beam is added to the pure VVB:
\begin{align}\label{eq:qplate_action_perturbed}
\hat Q_{\pi+2\varepsilon}\, \hbox{TEM}_{0,0}\,\ket{H}&=i\,\sin{(\pi/2+\varepsilon)}\,\ket{VVB}_1^{rad}+\cos{(\pi/2+\varepsilon)}\, \hbox{TEM}_{0,0}\,\ket{H}\simeq\nonumber\\
&\simeq i\,\left(\ket{VVB}_1^{rad}+i\,\varepsilon\, \hbox{TEM}_{0,0}\,\ket{H}\right).
\end{align} 
A similar expression holds for an azimuthal VVB. In close analogy to Eq.\ \ref{eq:vvb_perturb}, the latter equations show that a small variation of $\delta$ can be treated as a perturbation to the original VVB. In Fig.\ \ref{fig: V-stability} we show a simulation of the polarization distribution of perturbed radial and azimuthal VBs (Eq.\ \ref{eq:qplate_action_perturbed}). In particular in the upper part of panels {\bf a-b} we plot a 2D map of the orientation angle $\psi$ of the local polarization ellipse, calculated in terms of the reduced Stokes parameters; here the two $C$-points are clearly visible as vortices of this scalar field (the ellipse orientation), with their separation changing with the value of $\delta$.\\ 
\linebreak
{\bf Experimental results}
\begin{figure}[t!]
	\centering
\includegraphics[width=17 cm]{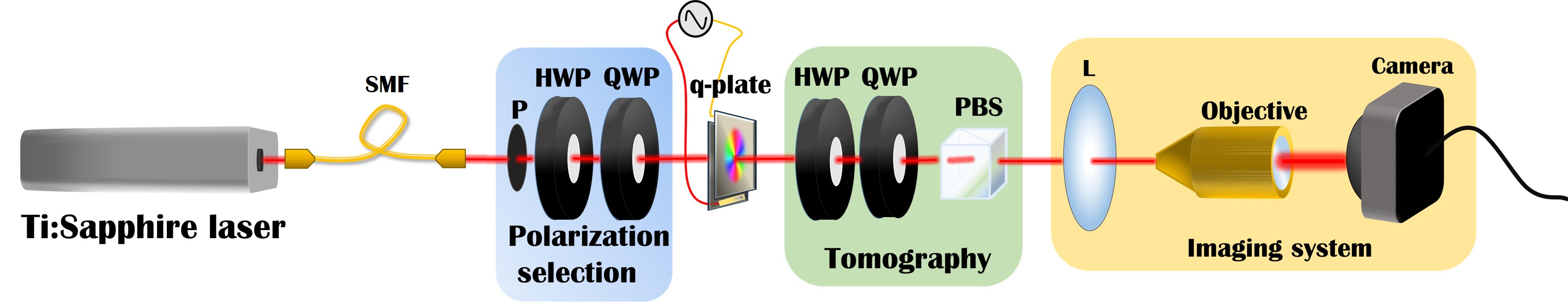}
	\caption{Experimental apparatus. A TEM$_{0,0} $ beam is obtained by filtering the output of a Ti:Sapphire laser through a single mode fiber (SMF). The initial polarization state is selected by using a polarizing beam splitter (PBS), a half-wave plate (HWP) and a quarter-wave plate (QWP). Then the beam passes through an electrically tunable $q$-plate, whose optical retardation is controlled by applying an adjustable electric field. When exiting the $q$-plate the beam has acquired an inhomogeneous polarization pattern that can be experimentally reconstructed by employing a point by point Stokes polarimetry, as discussed in the main text. The projection over the six polarization states $H$, $V$, $A$, $D$, $L$, $R$ is implemented through a QWP and a HWP followed by a PBS. The intensity of the analyzed field component is recorded on a CMOS camera. A lens (focal length $f=10 $ cm) and a 20X microscope objective placed on a translation stage are used to study the polarization pattern at different distances $z$ from the $q$-plate.}
	\label{fig: APPARATO}
	\end{figure}
To confirm the theoretical predictions discussed in the previous section, we implemented the setup shown in Fig.\ \ref{fig: APPARATO}. The output of a Ti:Sa laser (wavelength $\lambda$=800 nm) is coupled into a single-mode fiber (SMF), used as a spatial filter in order to produce a pure $\hbox{TEM}_{0,0}$ Gaussian mode at the input of the setup. At the exit of the SMF, the beam (uniform) polarization is prepared into vertical or horizontal states by means of a linear polarizer followed by a half-wave plate (HWP). A $q$-plate ($q$=1/2) with optical retardation $\delta$, whose value is controlled through a tunable electric field applied to the outer faces of the cell \cite{Piccirillo2010}, transforms the beam into the VB reported in Eq.\ \ref{eq:qplate_action_perturbed}. In order to reconstruct the 2D polarization pattern in a transverse plane we implemented a point-by-point polarization analysis, similar to that reported in Ref.\ \cite{Cardano2012}. For each beam configuration, on a CMOS camera (1280x1024 pixels) we recorded the intensity profile of the field components associated with $\{H,V\}$, $\{L,R\}$ and diagonal and anti-diagonal ($\{D,A\}$) polarization states. These components are selected by rotating suitably a set of waveplates, followed by a linear polarizer. By using a dedicated software, Stokes parameters are calculated point-by-point according to the definitions $S_0=I_H+I_V$, $S_1=I_H-I_V$, $S_2=I_D-I_A$, $S_3=I_L-I_R$; here $I_j$ represent the measured intensities of the six polarization components, with $j\in\{H,V,D,A,L,R\}$. To take into account small fluctuations of the beam position with respect to the camera the field intensities are averaged over arrays of 3x3 pixels. 
An imaging system made of a lens (focal length = 10 cm) followed by a microscope objective is used to determine the polarization pattern at different positions along the propagation axis $z$.\\
\begin{figure}[t!]
\centering
\includegraphics[width=17 cm]{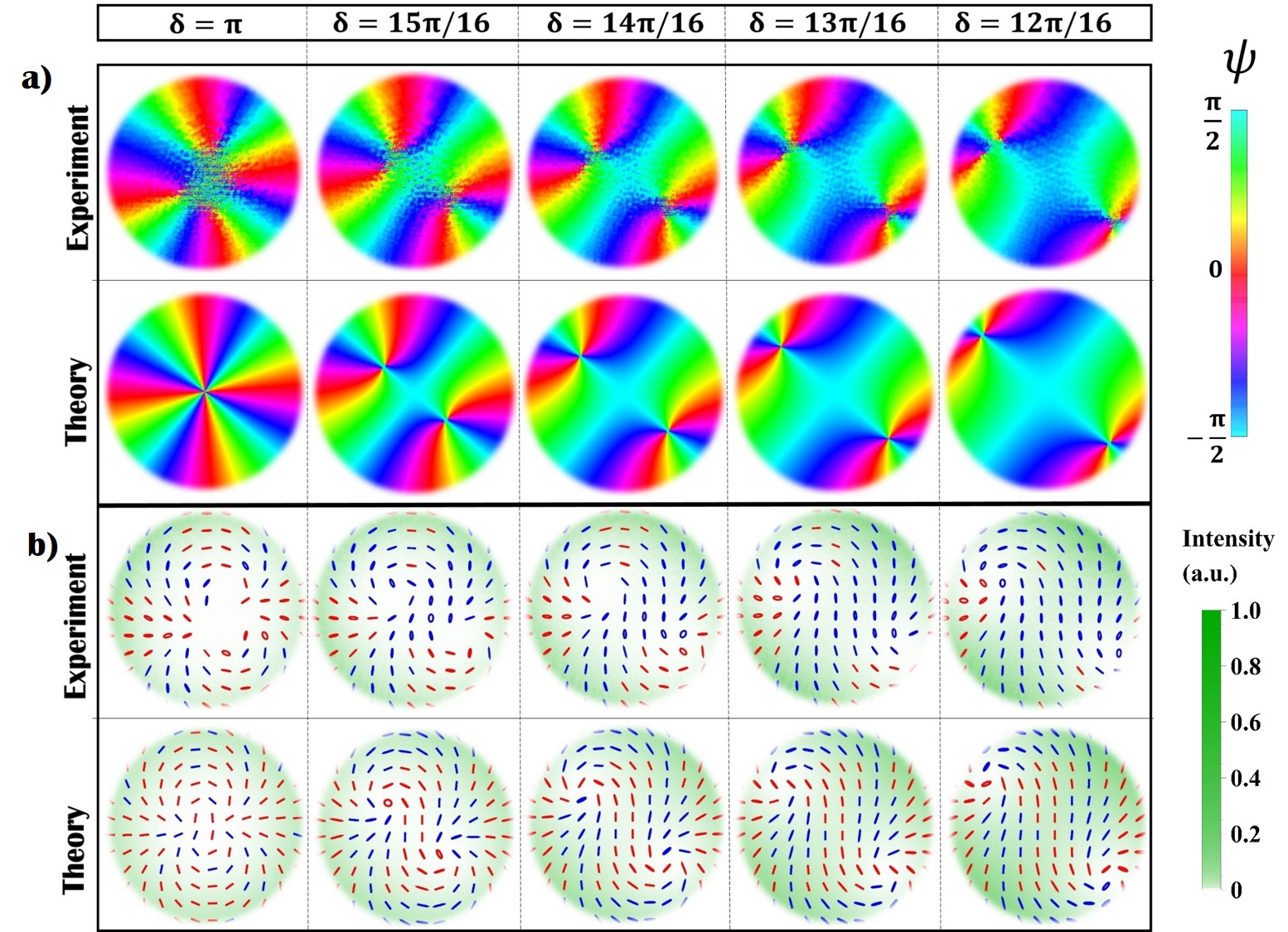}
\caption{Instability of a higher-order $V$-point in VVB with $\eta=2$ generated by a $q$-plate with $q=1$ and $\alpha_0=2.1$ rad, acting on a TEM$_{0,0}$ vertically polarized beam. a) We plot the orientation angle $\psi$ of the measured polarization ellipses, and the corresponding theoretical predictions, when varying $\delta$ between $\delta=\pi$ and $\delta=3\pi/4$ with steps of $\pi/16$. Simulations show that the original $V$-point splits into two pairs of $C$-points; similarly to the previous case, in each pair the two singularities have opposite handedness. In the experimental data, the formation of these two pairs can be observed clearly: as discussed in the main text, this is associated with the decay of the high-order phase vortex in each of the two circular components. Nevertheless, for each pair the system spatial resolution does not allow to distinguish two different $C$-points, because they remain very close to each other, at least for small deviations from the ideal case. In panel b, we show the corresponding experimental and theoretical polarization patterns associated with these fields. Here, red and blue colored  ellipses are associated with left ($s_3>1$) and right handed ($s_3<1$) polarization states, respectively.
}
\label{fig: V-stability-q1}
\end{figure}
By introducing a tiny alteration of the $q$-plate voltage with respect to the optimal condition $\delta=\pi$, we investigated the instability of a $V$-point singularity that transforms into a pair of $C$-points. When $\delta=\pi$, the polarization is linear in every point of the transverse plane and has a radial or azimuthal pattern, depending on the input polarization. If we introduce a small detuning, that is $\delta\rightarrow\pi-2\epsilon$, a uniform polarized Gaussian beam is added coherently to the original VVB. As previously discussed, $2\eta$ $C$-points are expected to form in place of the original singularity with topological charge $\eta$. In Fig.\ \ref{fig: V-stability} we show the experimental results that confirm these predictions. In order to unveil the formation of $C$-points pairs, that occurs very close to the beam center, we used a single lens to image this small portion of the beam on the camera sensor. We imaged on the camera the beam at $z=0.22\,z_R$ and considered only the region $r<0.58\,w_0$, where $z_R$ and $w_0$ are the Rayleigh range and the beam waist, respectively. The measured Stokes parameters are used to determine the polarization pattern and, as a consequence, the orientation angle $\psi$ of the local polarization ellipses, calculated as $\psi=\frac{1}{2}\arg{\left(S_1+i\,S_2\right)}$. In good agreement with the theoretical predictions (see Fig.\ \ref{fig: V-stability}c-d), we observe the original $V$-point splitting into two $C$-points with opposite handedness, whose spatial separation grows as $\epsilon$ is increased, in agreement with Eq.\ \ref{eq:cpoints}. The topological charge of such $C$-points is $1/2$, hence the total charge is conserved.\\

\begin{figure}[b!]
\centering
\includegraphics[width=17 cm]{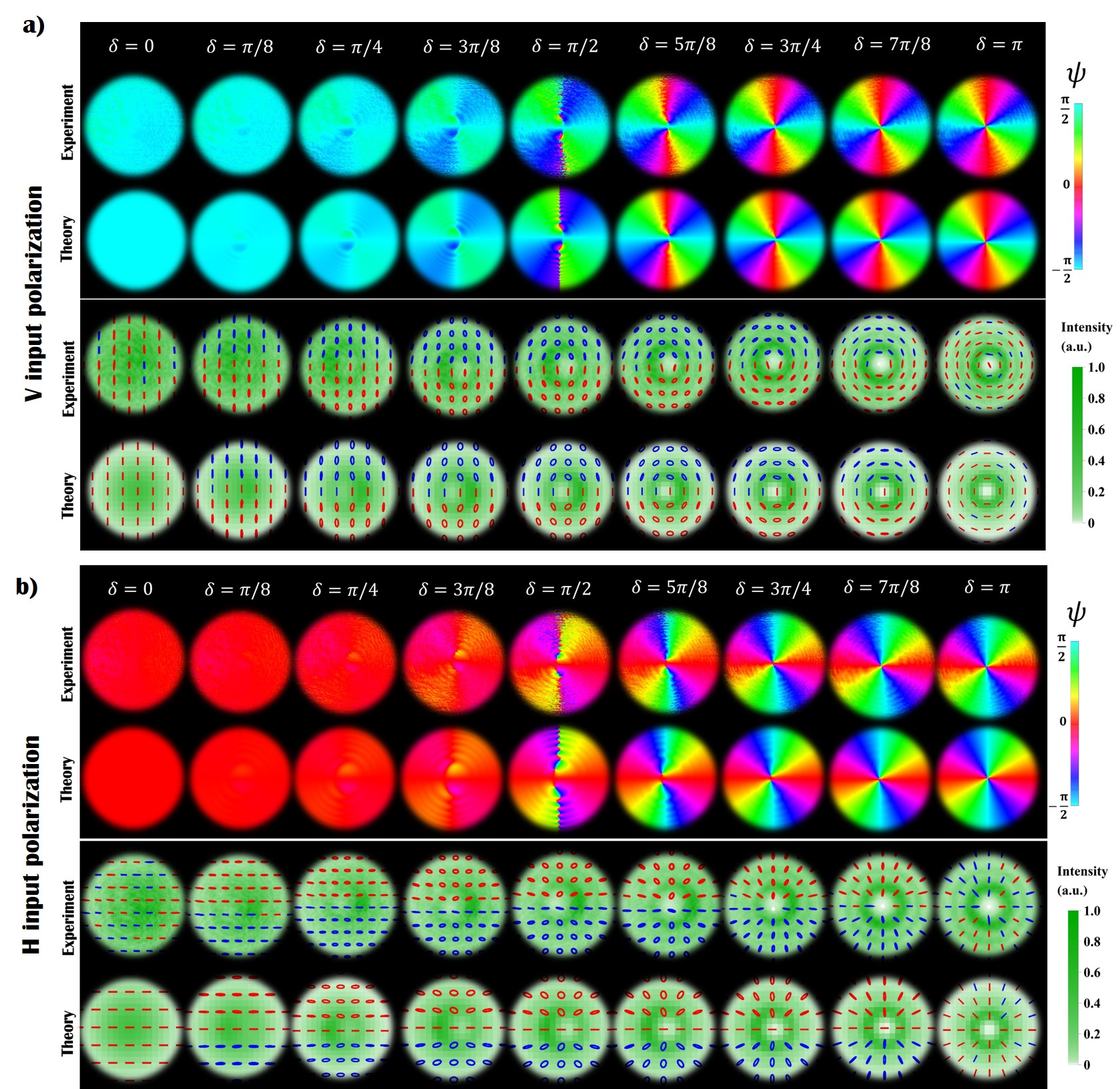}
\caption{Transition from a trivial to a topologically non-trivial polarization pattern. In panels a-b we report the experimental and theoretical intensity and polarization distribution of the near field ($\zeta=0.01$) for both radial and azimuthal VVB, respectively. We observe no polarization singularities (actually, $C$-points exist in a region where the beam intensity is too low, as shown in Fig.\ \ref{fig: phasetransition2}) when $\delta<\pi/2$. At $\delta=\pi/2$ an undefined number of $C$-point dipoles appears, in proximity of intersections between the Gaussian and the oscillating HyGG$_{-1,1}$ envelopes. When $\delta$ is increased, a pair of $C$-points with $\eta=1/2$ appears in the polarization pattern, with the distance between these points getting smaller as $\delta\rightarrow\pi$. The topological charge associated with a path enclosing both singularities is $\eta=1$. Here, red and blue colored  ellipses are associated with left ($s_3>1$) and right handed ($s_3<1$) polarization states, respectively.}
\label{fig: v_phase_topological_transition}
\end{figure}
\begin{figure}[]
\centering
\includegraphics[width=17 cm]{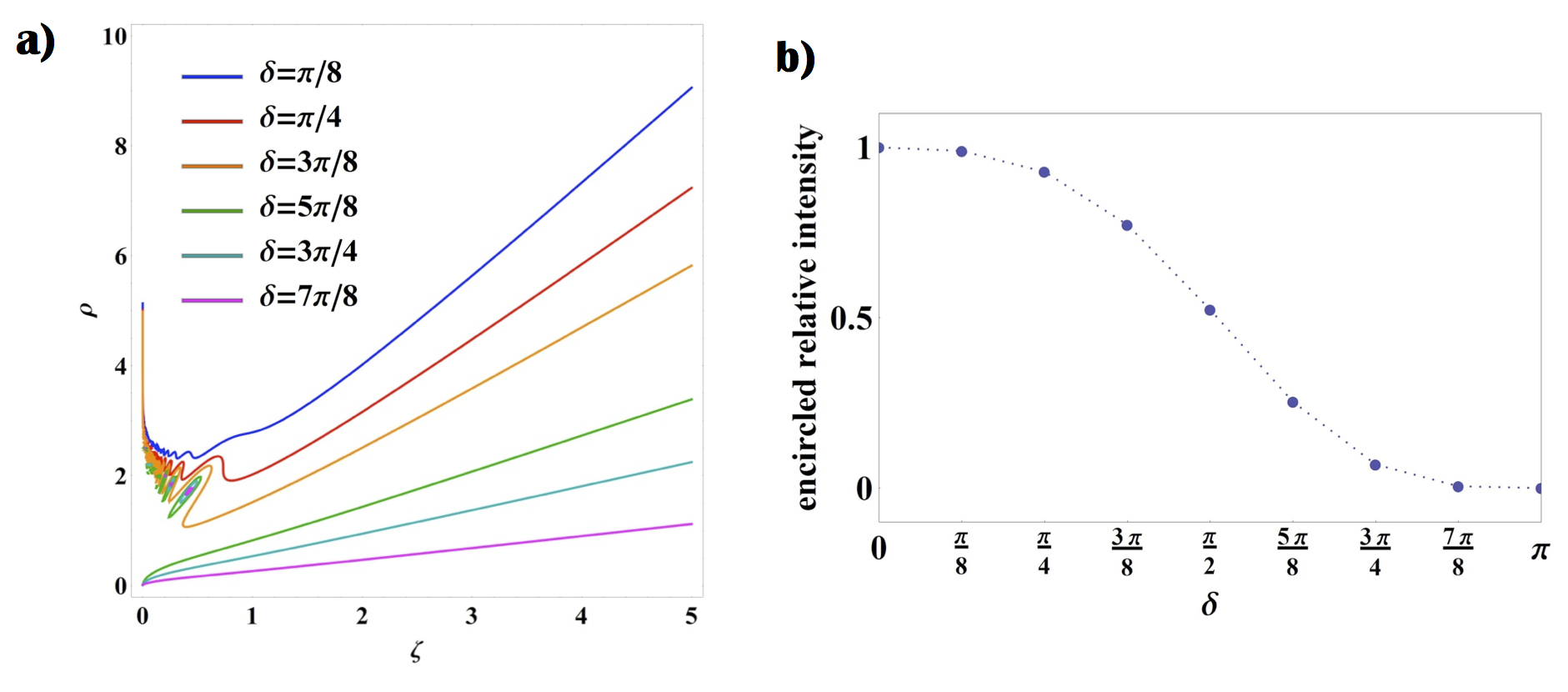}
\caption{Dynamical evolution of $C$-points during propagation and associated enclosed energy. a) We plot the radial coordinate $\rho$ of the position of left handed $C$-points vs the longitudinal coordinate $\zeta$, for different values of $\delta$ (the latter are encoded in the color of the curve, as displayed in the figure legend). Two regimes can be distinguished: for $\pi/2<\delta<\pi$ $C$-points merge at the beam center ($\rho=0$) as $\zeta\rightarrow0$, hence making the polarization pattern topology non-trivial. In the same regime, additional closed loops indicate the existence of $C$-point dipoles appearing and disappearing as the beam propagates, as a consequence of the dynamical evolution of the radial ripples characterizing HyGG modes. For $0<\delta<\pi/2$, instead, the $C$-point radius increases indefinitely as we approach the beam near field. b) For the same values of $\delta$ as in panel a, we plot here the fraction of the beam intensity contained in a circular region with a radius given by the $C$-point radial coordinate, in the far field limit. These results, in particular, are obtained when considering $\zeta=15$, but in the limit of large $\zeta$, they remain essentially constant.}
\label{fig: phasetransition2}
\end{figure}
\begin{figure}[t!]
\centering
\includegraphics[width=17 cm]{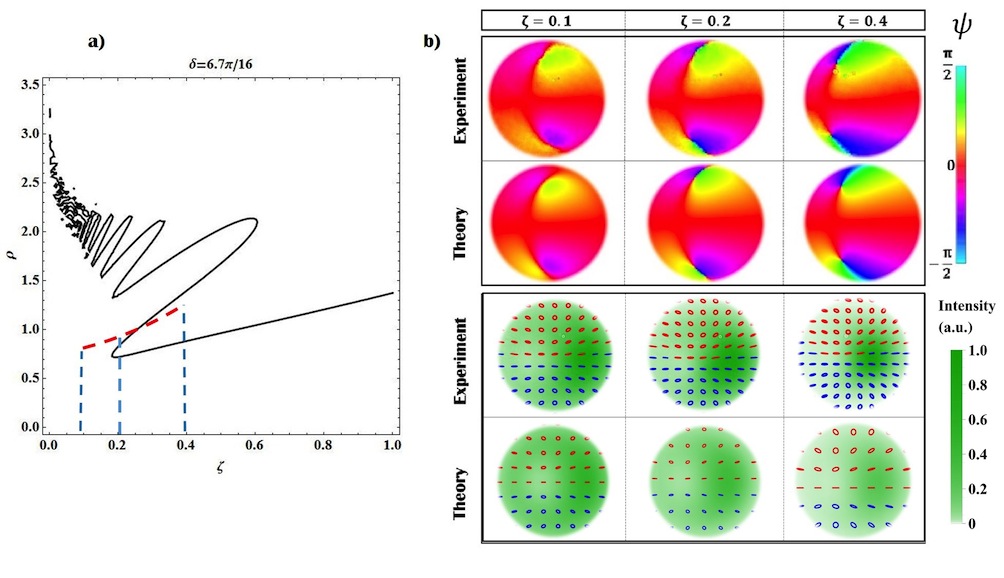}
\caption{Dynamical formation of $C$-points. Here we show that, for $\delta<\pi/2$, $C$-points (absent in the near field) appear during propagation. In panel a) we report the $C$-point position versus the longitudinal coordinate $\zeta$ in the case $\delta=6.7\pi/16$, for a $H$ polarized input beam. The red line indicates the dimension of the regions investigated in the experiment while blue dashed lines show the corresponding values of $\zeta$. In panel b), we report the plot of the orientation angle $\psi$ and the polarization pattern of the associated beam; here, theoretical simulations are added for comparison. Data refers to three different propagation distances, as reported in the plot: when $\zeta=0.1$, no singularities are visible in the polarization pattern; at $\zeta=0.2$, two $C$-point dipoles are clearly emerging, and a single pair is observed at $\zeta=0.4$. In the polarization plots, red and blue colored  ellipses are associated with left ($s_3>1$) and right handed ($s_3<1$) polarization states, respectively.}
\label{fig: farfield}
\end{figure}

The same phenomenon can be observed for $V$-points with higher topological charge, obtained using $q$-plates with $|q|>1/2$. In Fig.\ \ref{fig: V-stability-q1} we plot the polarization pattern and the orientation angle of $\psi$ measured for a $q$-plate with $q=1$ and $\alpha_0=3\pi/4$. Here a $V$-point with charge $\eta=2$ is observed to split into four $C$-points. For each circular component, the central vortex has a charge $\pm2$; as mentioned previously these are unstable and decay into two equally charged vortices \cite{Freund1999,Ricci2012}; this process is much faster (with respect to a variation of $\delta$) if compared to the $V$-point splitting discussed previously (see Fig.\ \ref{fig: V-stability}). Two pairs of $C$-point move away from the beam center as $\delta$ decreases, although in each pair the distance between the two singularities remains small so that they cannot be clearly resolved in our system.\\
By changing the voltage applied to the $q$-plate we can tune the device retardation to any value in the range $(0,2\pi)$, thus we can also explore what happens when the Gaussian term becomes comparable to the VVB amplitude. In Fig.\ \ref{fig: v_phase_topological_transition} we show the   polarization and intensity patterns measured in the near field of the beam (compared with theoretical predictions), obtained when varying $\delta$ between 0 and $\pi$ with steps of $\pi/8$. Theoretical simulations are added here for comparison. At a glance, when decreasing $\delta$, $C$-points are observed to move away from the beam center and seem to disappear when $\delta<\pi/2$. Accordingly, the topological features of the polarization pattern change abruptly when the amplitude of the Gaussian term becomes higher than the original VVB. This is not surprising, as we are exploring an intermediate regime between the extreme cases $\delta=0$ (a Gaussian beam with no polarization singularities) and $\delta=\pi$ (VVB beam with a $\eta=1$ $V$-point), which have different topological features. However, as the VVB and the perturbing term diffract differently, these features are expected to change when the beam propagates, making the situation much more complex. Importantly, this behavior will depend also on the VVBs radial profile, here assumed to be that of HyGG modes; although the generalization to other types of helical modes is out of the scope of this work, in the Methods we briefly discuss the simple case in which the VVBs radial profile is that of lowest radial-order Laguerre-Gauss modes with $p=0$. In this specific case indeed an analytical expression for the $C$-point positions can be found. For our specific configuration, we investigate the dynamical evolution of polarization singularities by considering the expression of the beam generated by a $q$-plate (with $\alpha_0=0$) when shined with a $H$ or $V$ polarized Gaussian beam:
\begin{equation}
\begin{split}\label{eq:vvb_OUTexpr} 
\ket{OUT}=&\biggr(\hbox{TEM}_{0,0}(\rho,\zeta)\cos{\left(\delta/2\right)}\pm i\hg{-|m|,|m|}(\rho,\zeta)e^{im\phi}\sin{\left(\delta/2\right)}\biggr)\ket{R}\\ &+ \biggr(\pm\hbox{TEM}_{0,0}(\rho,\zeta)\cos{\left(\delta/2\right)}+i\hg{-|m|,|m|}(\rho,\zeta)e^{-im\phi}\sin{\left(\delta/2\right)}\biggr)\ket{L},
\end{split}
\end{equation}
where the $\pm$ sign stands for $H$ or $V$ input polarization, respectively, and $m=2q$. \hbox{Dimensionless} units are introduced here, where the distance from the $q$-plate $z$ and the radial coordinate $r$ are normalized with respect to the Rayleigh range $z_R$ and the waist $w_0$ of the beam, respectively ($\zeta=z/z_R$, $\rho=r/w_0$). $C$-point positions are calculated by solving the equations $C_{L/R}(\rho,\phi,\zeta)=0$, where left and circular components $C_{L/R}$ are those reported in Eq.\ \ref{eq:vvb_OUTexpr} (see the Methods for details about the solution of such equations). In Fig.\ \ref{fig: phasetransition2}a we plot the position of left-handed $C$-points as a function of the coordinate $\zeta$, for the case $m=1$ (similar features are obtained for higher values of $m$). For small values of $\zeta$, multiple rings characterize the radial distribution of HyperGeometric-Gaussian modes and, as a result, many $C$-points may appear at given transverse plane. However, for any value of $\delta$ and $\zeta$, there exist at least a left-handed $C$-point (and its right-handed partner), although it might be positioned in the peripheral regions of the beam where the field intensity is negligible. For high values of $\zeta$ (far field), numerical simulations show that the distance between $C$-point positions and the beam axis is proportional to the coordinate $\zeta$. In this configuration, we can evaluate if these singularities can be still considered within the beam; in particular, for different values of $\delta$ we can compute the fraction of the beam intensity contained in the circular region delimited by the $C$-point radial position. In Fig.\ \ref{fig: phasetransition2}b we plot the relative encircled intensity of the beam, which is observed to increase as the Gaussian contribution becomes stronger. When $\delta<\pi/8$, for example, more than the 99\% of the beam intensity is contained in the radius defined by the $C$-point position, i.e. the singularities can be considered as lying outside the beam.
In the near field, instead, qualitative differences manifest depending on the $q$-plate retardation being higher or lower than $\pi/2$. In particular, in the latter case the singularities move away at infinite distances, while in the former, $L$ and $R$ $C$-points merge at the beam center (forming a single $V$-point). These different features are a consequence of the irregular behavior of HyGG$_{-|m|,m}$ modes in the limit $\zeta\rightarrow 0$. As discussed in the Methods indeed (see Eq.\ \ref{eq:Cpointcurve}) $C$-points can form at the intersections between the Gaussian and HyGG$_{-|m|,m}$ envelopes; when $\delta > \pi/2$, these envelopes cross in the region of the central dip characterizing HyGG$_{-|m|,m}$ modes, at a specific radial distance that depends on $\delta$. However, since the dip width vanishes when approaching the near field, independently of $\delta$ all $C$-points merge at the center. When $\delta<\pi/2$, singularities do not form in the HyGG dip but rather at the intersection of the tails of both Gaussian and HyGG modes; such intersections exist since the latter have a larger radial profile, for finite values of $\zeta$. But as $\zeta\rightarrow0$, the HyGGs envelope converges to that of the Gaussian beam, and clearly two Gaussian profiles with different amplitudes cannot have intersection points, so that all $C$-points are expelled from the beam. In the Methods, we repeat this analysis for the simpler case in which Laguerre-Gauss modes are taken in the place of HyGG modes. In such configuration indeed an expression for $C$-points position can be derived analytically.\\
As shown in Fig.\ \ref{fig: phasetransition2}a, VBs obtained when $\delta<\pi/2$ may show $C$-points only after a definite value of $\zeta$. We confirm experimentally this effect by investigating the dynamical formation of such singular points in the polarization pattern of a beam obtained when $\delta=6.7\pi/16$. In Fig.\ \ref{fig: farfield} we report the experimental data and the associated theoretical predictions. Although no $C$-points are observed in the near field (see Fig.\ \ref{fig: v_phase_topological_transition}), they appear as we increase the propagation distance $\zeta$, in agreement with our previous discussion (see Fig.\ \ref{fig: phasetransition2}a). In particular, at $\zeta=0.2$ we can observe a double pair of singularities, as a consequence of the oscillatory behavior of the VVB amplitude profile.\\
\section*{Discussion}
In this study we investigated the topological features of vector vortex beams and the robustness of the associated singularities when introducing a perturbation to the field. Polarization singularities manifesting at the center of such beams are unstable and transform into multiple $C$-points with equal topological charge $\pm 1/2$, the lowest order singularities of fully polarized light. Here we perturb a VVB by adding coherently a tunable amount of a linearly polarized Gaussian beam and demonstrate experimentally a possible mechanism that leads to the unfolding of the central singularity, in analogy to similar phenomena observed in the skylight polarization \cite{Berry2004} or in high-order optical vortices \cite{Dennis2009}. On one hand, this realization provides a simple example of transformations between different polarization singularities \cite{Bliokh2008,Flossmann2005,Freund2002,Vyas}; on the other, it allows for a detailed investigation of phenomena that may affect optical systems exploiting VVBs, for which the presence of the fundamental TEM$_{0,0}$ mode can result from different types of misalignments \cite{Kumar2011,Dennis2006,Bekshaev2004}, scattering \cite{Ricci2012} or from turbulence in the propagation medium \cite{Cheng2009}. Investigating the stability of VVBs can be of interest for all photonic applications involving these structured beams, since a modification of the intensity pattern always accompanies the singularity splitting. As discussed before, similar alterations of a VVB can occur as a consequence of experimental imperfections, and the beam distortion caused by the $V$-point splitting may play a role in a variety of applications, as for example in material shaping \cite{Nivas2015}. Reversing the current approach, as in prospect it could be possible to tune the $q$-plate optical retardation in order to compensate the effect of experimental imperfections and reduce such deformations of the beam profile \cite{Neo2014}. Finally, our results may find application in the context of singularimetry; weak fields can indeed be measured by letting them perturb unstable optical fields, and features of materials that have interacted with such beams can be extracted from the pattern formed by split singularities \cite{Dennis2012}. In addition, since perturbations can be introduced by imperfections in the optical setup, the unfolding of the central $V$-points can be used to assess the quality of the VVB generation system \cite{Ricci2012}.  

\begin{footnotesize}

\section*{Methods}

{\bf Helical modes of light}\label{app:B}
Laguerre-Gaussian and Hypergeometric-Gaussian beams \cite{Karimi2007} represent specific cases of the so-called Circular Beams (CB) \cite{Vallone2015}, a class of optical spatial modes characterized by the phase factor $\exp(im\phi)$ associated with the orbital angular momentum. As in the main text, we used adimensional cylindrical coordinates $\rho=r/w_0$ and $\zeta=z/z_R$, where $w_0$ is the waist radius of the Gaussian envelope and $z_R$ the Rayleygh range, respectively. \\
Laguerre-Gaussian $LG_{p,m}$ modes have the well known expression:
\begin{equation}
\begin{split}\label{eq:LGmodes}
LG_{p,m}(\rho,\zeta,\phi)=&\sqrt{\frac{2^{|m|+1} p!}{\pi (p+|m|)!(1+\zeta^2)}}\Biggr(\frac{\rho}{\sqrt{1+\zeta^2}}\Biggr)^{|m|}\exp\biggr(-\frac{\rho^2}{1+\zeta^2}\biggr){L_p}^{|m|}(2\rho^2/(1+\zeta^2))\times \\
&\exp\biggr(i\frac{\rho^2}{\zeta+1/\zeta}\biggr)\exp[im\phi-i(2p+|m|+1)\arctan(\zeta)],
\end{split}
\end{equation}
where $L_{p}^{|m|}(x)$ is the generalized Laguerre polinomial and $p$ is a positive integer.\\
A $q$-plate shined by a TEM$_{0,0}$ mode generates the so-called Hypergeometric-Gaussian modes \cite{Karimi2009} (see Eq.\ \ref{eq:qplate_action}):
\begin{equation}
\begin{split}\label{eq:hgmodes}
\hg{p,m}(\rho,\zeta,\phi)=&\sqrt{\frac{2^{1+|m|+p}}{\pi\Gamma(1+|m|+p)}}i^{|m|+1}\frac{\Gamma(1+|m|+p/2)}{\Gamma(1+|m|)}\zeta^{p/2}(\zeta+i)^{-(1+|m|+p/2)}\rho^{|m|}\times \\
&\exp(-i\rho^2/(\zeta+i)+im\phi) _{1}F_{1}\biggr(-p/2;|m|+1;\rho^2/(\zeta(\zeta+i))\biggr),
\end{split}
\end{equation} 
where $\Gamma(z)$ is the Euler Gamma function and $F_{1}(a;b;z)$ is the confluent Hypergeometric function. Through the main text we used the notation $\hg{p,m}$ when referring only to the radial distribution of the associated HyGG modes, not including the azimuthal phase factor $e^{i m\phi}$.\\
\linebreak
{\bf Determination of $C$-point position}\label{app:A}
Here we give a detailed description of the derivation of the $C$-point position (see Fig.\ \ref{fig: phasetransition2}a). We recall here the general expression describing a beam generated by a $q$-plate with $\alpha_0=0$ when shined by a $H$ or $V$ polarized Gaussian beam (already reported in Eq.\ \ref{eq:vvb_OUTexpr}):
\begin{equation}
\begin{split}\label{eq:vvb_OUTexpr1} 
\ket{OUT}=&C_R(\rho,\zeta,\phi;\delta)\ket{R}\pm C_L(\rho,\zeta,\phi;\delta)\ket{L}\\=&\biggr(\hbox{TEM}_{0,0}(\rho,\zeta)\cos{\left(\delta/2\right)}\pm i f_{|m|}(\rho,\zeta)e^{-im\phi}\sin{\left(\delta/2\right)}\biggr)\ket{R}\\ & + \biggr(\pm\hbox{TEM}_{0,0}(\rho,\zeta)\cos{\left(\delta/2\right)}+if_{|m|}(\rho,\zeta)e^{im\phi}\sin{\left(\delta/2\right)}\biggr)\ket{L},
\end{split}
\end{equation}
where the plus or minus sign is for $H$ and $V$ input polarizations, respectively, $\zeta=z/z_R$ is the propagation distance normalized with respect to the Rayleigh range $z_R$, and $m=2q$.\\
In our experiment the function $f_{|m|}$ is given by Hypergeometric Gaussian mode $\hg{-|m|,m}$. It is worth noting that specific architectures allow using the $q$-plate to generate helical modes with a different radial profile; as an example, recently a $q$-plate placed inside a laser cavity has been exploited for the generation of high quality Laguerre-Gauss VBs \cite{Naidoo2015} with $p=0$. For this reason, we consider here also the case $f_{|m|}=LG_{0,m}$. As we will show in the following, the $C$-points positions can be deduced analytically in this case. \\
The $C$-point position at a given $\delta$ and $\zeta$ can be obtained simply by solving the implicit equation  $C_{L,R}(\rho,\zeta,\phi;\delta)=0$ (see Eq.\ \ref{eq:vvb_OUTexpr}). We limit ourselves to searching for the distance of $C$-points from the center, which can be found by solving the simplified equation $|C_{L,R}(\rho,\zeta,\phi;\delta)|^2=0$. Explicitly, this reads:
\begin{equation}
\begin{split}\label{eq:cpointcurve}
&\cos^2{\left(\delta/2\right)}|\hbox{TEM}_{0,0}(\rho,\zeta)|^2+\sin^2{\left(\delta/2\right)}|f_{|m|}(\rho,\zeta)|^2\\& \pm 2\cos{\left(\delta/2\right)}\sin{\left(\delta/2\right)}Re\{ie^{-im\phi}\hbox{TEM}^{*}_{0,0}(\rho,\zeta)f_{|m|}(\rho,\zeta)\}=0.
\end{split}
\end{equation}
A solution for such equation exists only if the following condition holds
\begin{equation}\label{eq:Cpointcurve}
\cos^2{\left(\delta/2\right)}|\hbox{TEM}_{0,0}(\rho,\zeta)|^2=\sin^2{\left(\delta/2\right)}|f_{|m|}(\rho,\zeta)|^2;
\end{equation}
hence, by solving jointly Eq.\ \ref{eq:cpointcurve}-\ref{eq:Cpointcurve} we are left with an implicit equation for $\rho$ as function of $\zeta$. If needed, then the solution can be inserted into Eq.\ \ref{eq:cpointcurve} to find the azimuthal coordinates of the singularities. \\
The case $f_{|m|}=\hg{-|m|,|m|}$ can be solved only numerically. Some solutions are shown in Fig.\ \ref{fig: phasetransition2}a and discussed in the main text. Here we focus on the case $f_{|m|}=LG_{0,m}$ where we can find an analytical expression for $C$-points positions as a function of $\zeta$. Importantly, this kind of beams (VVBs whose radial distribution is that of LG modes with the lowest radial index $p=0$) can also be generated experimentally, as for example by exploiting a $q$-plate placed inside a laser cavity \cite{Naidoo2015}. In this case Eq.\ \ref{eq:Cpointcurve} reads:
\begin{equation}
\Biggr|\frac{LG_{0,m}(\rho,\zeta)}{TEM_{0,0}(\rho,\zeta)}\tan(\delta/2)\Biggr|^2=\frac{2^{|m|}}{m!}\frac{\rho^{2|m|}}{(1+\zeta^2)^{|m|}}\tan^2(\delta/2)=1.
\end{equation}
It follows that the distance of $C$-points from the beam center $\rho_C(\zeta)$ is given by:
\begin{equation}\label{eq:cpoint_position}
\rho_C(\zeta)=\Biggr[\sqrt{\frac{m!}{2^{|m|}}}\cot^2(\delta/2)\Biggr]^{1/2|m|}\sqrt{1+\zeta^2}.
\end{equation}
In order to evaluate if such singularities are contained in the beam or not, one can either apply the same approach used for the case of HyGG modes (see Fig.\ \ref{fig: phasetransition2}b), or directly compare the position of singular points with the beam radius. This is typically defined as the root-mean-square $\sigma_{rms}$ of the beam intensity in the transverse plane \cite{Vallone2016a}:
\begin{equation}
\sigma_{rms}^2(\zeta)=\int I(\rho,\zeta,\phi)\rho^3 d\rho d\phi,
\end{equation}
where $I(\rho,\zeta,\phi)$ is the beam intensity. From Eq.\ \ref{eq:vvb_OUTexpr1} with $f_{|m|}=LG_{0,m}$ we obtain:
\begin{equation}\label{eq:beamradius}
\sigma_{rms}(\zeta)=\sqrt{\frac{1+|m|\sin^2(\delta/2)}{2}}\sqrt{1+\zeta^2}.
\end{equation} 
In the far field, the corresponding beam divergence is given by $\theta_{rms}=\sigma_{rms}/\zeta$, in the limit $\zeta\rightarrow\infty$. The associated expression is simply: 
\begin{align}\label{eq:thetarms}
\theta_{rms}(\delta,|m|)=\sqrt{[1+|m|\sin^2(\delta/2)]/2},
\end{align}
By comparing Eq.\ \ref{eq:beamradius} and Eq.\ \ref{eq:cpoint_position} we can observe that the beam radius has the same functional form as the $C$-point position. The divergence of the latter is: 
\begin{align}\label{eq:thetaC}
\theta_c(\delta,|m|)=\Biggr[\sqrt{\frac{m!}{2^{|m|}}}\cot^2(\delta/2)\Biggr]^{1/2|m|}.
\end{align} 
By comparing Eq.\ \ref{eq:thetarms} and Eq.\ \ref{eq:thetaC} it is possible to compute exactly whether $C$-points diverge more or less rapidly than the beam radius, for any value  of $\delta$. However, we observe that our analysis aims only at identifying qualitative features: quantitative results are ambiguous as multiple definitions of the beam divergence can be given (for instance the quantity $\sqrt{2}\theta_{rms}$ is often used). In the near field, at odds with the HyGG case, $C$-points have a definite position given by Eq.\ \ref{eq:cpoint_position} when $\zeta=0$. By comparing the latter with Eq.\ \ref{eq:beamradius} it is possible to check if such singular regions are contained inside the beam. In conclusion, we point out that this analysis relying on the r.m.s. as a measure of the beam width cannot be done in the case of $\hg{}$ beams; indeed, $|\hg{-|m|,m}|^2\propto r^{-4}$ when $r\rightarrow\infty$ (Ref.~\citenum{Karimi2007}) and the r.m.s becomes infinite for any value of $\zeta$.

\end{footnotesize}

%

\section*{Acknowledgements}
We thank Mark Dennis for preliminary reading of the manuscript and for providing useful comments. This work was supported by the European Research Council (ERC), under grant no.694683 (PHOSPhOR).
\section*{Author contributions statement}
A.D., M.M., F.C. and L.M. devised various aspects of the project and designed the experimental methodology. A.D., M.M. and F.C., with contributions from C.d.L., carried out the experiment and analyzed the data. B.P. prepared the q-plates. A.D. and F.C. wrote the manuscript, with contributions from M.M. and L.M.. All authors discussed the results and contributed to refining the manuscript.
\section*{Competing financial interest}
The authors declare no competing financial interests.
\end{document}